\documentclass[3p,11pt]{elsarticle}
\usepackage{amsmath,amssymb,epsfig,color,mathrsfs,bbm,subcaption}
\usepackage{graphicx}

\begin{document}

\title{Mode-by-mode hydrodynamics: ideas and concepts}

\author{Stefan Floerchinger}

\address{Physics Department, Theory Unit, CERN, CH-1211 Gen\`eve 23, Switzerland}

\begin{abstract}
The main ideas, technical concepts and perspectives for a mode resolved description of the hydrodynamical regime of relativistic heavy ion collisions are discussed. A background-fluctuation splitting and a Bessel-Fourier expansion for the fluctuating part of the hydrodynamical fields allows for a complete characterization of initial conditions,  the fluid dynamical propagation of single modes, the study of interaction effects between modes, the determination of the associated particle spectra and the generalization of the whole program to event-by-event correlations and probability distributions.
\end{abstract}

\maketitle

In recent years it has become apparent that what one may call the ``fluid dynamic standard model of heavy ion collisions'' according to which initial density anisotropies are evolved to final momentum anisotropies by almost ideal hydrodynamics works surprisingly well, see \cite{Heinz:2013th,Gale:2013da,Hippolyte:2012yu,Teaney:2009qa} for recent reviews. Some of the remaining puzzles are about the nature of the initial state directly after the collision and the early out-of-equilibrium dynamics that drives it towards local equilibrium on a rather short time scale, see \cite{Berges:2012ks} for an overview over recent literature on that question. 
On a more quantitative level one would like to better understand how thermodynamic and transport properties govern the hydrodynamic regime in order to allow for precise comparison between theoretical calculations and experimental measurements of these material properties of strongly interacting quantum field theory.

Interesting new insights into both the properties of the initial state and  the material properties of QCD in the hydrodynamic regime may come from the study of fluctuations in the hydrodynamic fields. More specific, from many models or on general grounds one expects event-by-event fluctuations around the average in the hydrodynamical fields such as energy density $\epsilon$, fluid velocity $u^\mu$, shear stress $\pi^{\mu\nu}$ (and more general also baryon number density $n_B$, electric charge density, electromagnetic fields and so on) at the initialization time $\tau_0$ where the hydrodynamical description becomes (approximately) valid.
These fluctuations should contain interesting information from early times, both from the initial state and the early non-equilibrium dynamics. Their dynamical evolution is governed by universal fluid dynamic equations, depending only on the thermodynamic and transport properties. Since one can distinguish fluctuations of different characteristic spatial size by their wave numbers, the information content is much richer than for averaged quantities only.

There is an interesting analogy to cosmology. Indeed, the cosmic microwave background also contains interesting information from early times and the time evolution is sensitive to the history of the universe. The spectrum of fluctuations involves many numbers (or whole functions) that can be compared between theory and experiment. Historically, the detailed study of this spectrum has lead to quite a detailed quantitative understanding of cosmology and one may hope that a similar progress is possible in heavy ion physics, as well.

So what would be the elements of a theoretical program that aims for a precise understanding of event-by-event hydrodynamical fluctuations in analogy to the corresponding program in cosmology? One may split this into four points:
\begin{enumerate}
\item Initial fluctuations at the initialization time of the hydrodynamic description have to be characterized and quantified completely. Ideally this is done in a way that is independent of specific models of the initial state. 
\item Fluctuations have to be propagated through the hydrodynamical regime and one would like to know how precisely this evolution is affected by the thermodynamical and transport properties (which are functions of temperature and chemical potential).
\item The impact of the fluctuations in hydrodynamical fields onto the particle spectra generated at freeze-out must be understood and quantified in detail.
\item The effects of fluctuations generated from non-hydro sources (such as for example jets) should be quantified and eventually taken into account, as well.
\end{enumerate}
We will go through these points again below. It is clear that they can be implemented in different ways in principle. Numerical event-by-event simulations for specific models of the initial state and specific choices of thermodynamic and transport properties implement already some parts of the above program. However, we will argue that a more analytic approach (based on a mode-decomposition) has certain advantages and in particular it allows to disentangle the different steps above. This will be useful in order to study more systematically what can be learned from experimental results, how they constrain models of the initial state (and early dynamics), how they constrain the thermodynamic and transport properties in the hydrodynamic regime and to what extent the two things are complementary.

Let us first concentrate on the initial state. For a particular event the initial conditions on the initialization surface (usually taken to be the one of fixed Bjorken time $\tau=\sqrt{t^2-z^2} = \tau_0$) are fixed in terms of a number of independent functions which one may choose to be enthalpy density $w$, three independent components of the fluid velocity $u^\mu$ and five independent components of the shear stress $\pi^{\mu\nu}$. (One may also take non-zero bulk viscous pressure, baryon number density, electric charge density etc. into account.)
One can always split these hydrodynamical fields into a background and a fluctuating part, for example
\begin{equation}
w = w_\text{BG} + \delta w,
\quad\quad
u^\mu = u^\mu_\text{BG} + \delta u^\mu.
\end{equation}
For the background one may use an event-averaged configuration which is as smooth and and symmetric as possible. In particular azimuthal rotation symmetry around the beam axis and Bjorken boost invariance in the longitudinal direction lead to useful simplifications. Note that one can also treat non-central collisions with a rotational symmetric background and in fact one can always consider an ensemble of events with arbitrary orientation in the transverse plane such that the event averaged hydrodynamic fields are rotational symmetric,
\begin{equation}
w_\text{BG} = \langle w \rangle=w_\text{BG}(\tau, r), \quad\quad u^\mu_\text{BG} = \langle u^\mu \rangle = (u^\tau_\text{BG}(\tau,r), u^r_\text{BG}(\tau,r),0,0).
\end{equation}
Due to the symmetries, the background enthalpy density and fluid velocity are determined by the two independent functions $w_\text{BG}(\tau_0, r)$ and $u^r(\tau_0,r)$ and similar for the shear stress. 

The fluctuation field or deviations from the background may not always be small. Independent of its size one can use a mode decomposition similar to a Fourier expansion. But what is the best way to do this expansion? In cosmology the rotational and translational symmetries of the background suggest for the fluctuations a decomposition into scalar-, vector- and tensor-perturbations which are subsequently expanded into conventional Fourier modes. The situation is more complicated in the heavy ion case since there is less amount of symmetry. 

For the azimuthal dependence it is clear that a Fourier expansion is the right thing to do and similar (but less common) for the rapidity dependence when the background is Bjorken boost symmetric. The radial dependence is more subtle. For transverse densities such as enthalpy density the characterization is traditionally done in terms of 
eccentricities $\epsilon_{n,m}$ and angles $\psi_{n,m}$ defined by
\begin{equation}
\epsilon_{n,m} \, e^{i m\, \psi_{n,m}}=\frac{\int dr\, \int_0^{2\pi}d\varphi \,r^{n+1}\, e^{im\varphi}\, w(r,\varphi)}{\int dr \,\int_0^{2\pi}d\varphi \,r^{n+1}\, w(r,\varphi)}. 
\end{equation}
In principle $w(r,\varphi)$ is completely determined by the set of all eccentricities $\epsilon_{n,m}$ and angles $\psi_{n,m}$. Closely related is the characterization based on cumulants \cite{Teaney:2010vd}. Although the characterization in terms of eccentricities or cumulants has been successful phenomenologically, there is, however, the drawback that no positive transverse density can be associated to a small set of eccentricities or cumulants (beyond Gaussian order) such that higher order eccentricities or cumulants vanish. Strictly speaking, this implies that single eccentricities (or a few of them) cannot be propagated hydrodynamically.
Moreover, the generalization of this kind of characterization to velocity and shear fluctuations is not known.

Another possibility to characterize a transverse density is in terms of orthonormal sets of functions. Different choices have been explored and used for different purposes \cite{Gubser:2010ui, Staig:2010pn, Florchinger:2011qf, ColemanSmith:2012ka}. Particularly useful is the following choice based on a Bessel-Fourier expansion and the background density \cite{Floerchinger:2013rya, Floerchinger:2013vua}
\begin{equation}
w(r,\varphi) = w_\text{BG}(r) + w_\text{BG}(r) \sum_{m=-m_\text{max}}^{m_\text{max}} \sum_{l=1}^{l_\text{max}} \tilde w^{(m)}_l \,e^{im\varphi} \;J_m(k^{(m)}_l r),
\label{eq:BesselFourierExpDensity}
\end{equation}
where $J_m(z)$ are Bessel functions of the second kind, $z^{(m)}_l$ their $l$'th zero crossings and
$k^{(m)}_l = z^{(m)}_l/R$. The radius $R$ is to some extend arbitrary but must be chosen large enough to the physical relevant region, for example $R=8$ fm does for LHC conditions.

Note that the density $w(r,\phi)$ is completely determined by set of all coefficients $\tilde w^{(m)}_l$ and that higher $l$ correspond to smaller spatial resolution. Also, due to the way the background density $w_\text{BG}(r)$ enters eq.\ \eqref{eq:BesselFourierExpDensity}, a single or a few non-zero coefficients $\tilde w^{(m)}_l$ lead to a positive density as long as the magnitude of the coefficients $|\tilde w^{(m)}_l|$ is not too large. This is not so for the closely related expansion proposed earlier in ref.\ \cite{ColemanSmith:2012ka}. It implies in particular that single (or a few) modes can be propagated in hydrodynamics without any further regularization or modification. 

Finally, a similar Bessel-Fourier mode expansion can be devised for vectors (velocity) and tensors (shear stress). For example, for the velocity one can write
\begin{equation}
\begin{split}
u^r = u^r_\text{BG} + \frac{1}{\sqrt{2}} (\tilde u^- + \tilde u^+), \quad\quad\quad u^\phi = \frac{i}{\sqrt{2} \,r} (\tilde u^- - \tilde u^+),
\end{split}
\end{equation}
with the two polarizations
\begin{equation}
\begin{split}
\tilde u^-(r,\phi) &= \sum_{m,l} \, \tilde u^{-(m)}_l \, e^{im\phi} \,J_{m-1}\left(k_l^{(m)}r \right),\\
\tilde u^+(r,\phi) &= \sum_{m,l} \, \tilde u^{+(m)}_l \, e^{im\phi} \,J_{m+1}\left(k_l^{(m)}r \right).
\end{split}
\end{equation}

The mode expansion is also useful to characterize whole event classes. This can be done in terms of a functional probability distribution $p_{\tau_0}[w, u^\mu, \pi^{\mu\nu}, \ldots]$. For the case where only the enthalpy density is allowed to fluctuate, $p_{\tau_0}$ becomes a function of the expansion coefficients $\tilde w^{(m)}_l$. For example, it may be of Gaussian form
\begin{equation}
p_{\tau_0} \sim  
\exp {\bigg [} - \frac{1}{2}\sum_{m=-m_\text{max}}^{m_\text{max}} \sum_{l_1,l_2=1}^{l_\text{max}} T^{(m)}_{l_1 l_2} \; \tilde w^{(m)*}_{l_1}   \tilde w^{(m)}_{l_2} {\bigg ]},
\end{equation}
in which case it is determined by correlator matrix
\begin{equation}
(T^{(m)})^{-1}_{l_1 l_2} = \langle  \tilde w^{(m)}_{l_1}  \tilde w^{(m)*}_{l_2} \rangle.
\end{equation}
More details on the Bessel-Fourier expansion of the hydrodynamic initial state and the characterization of event ensembles can be found in ref.\ \cite{Floerchinger:2013vua}. 

Now let us turn to step 2 in the above listing, the propagation of perturbations through the hydrodynamic regime. One has two options here. In principle one could solve the full hydro equations numerically for different combinations of initial modes and extract numerically the part of the solution that is linear, quadratic, cubic and so on in the initial perturbations.

It is more elegant, however, to derive evolution equations directly for the fluctuating part of the hydrodynamic fields by subtracting the background solution. For small enough fluctuations the resulting equations can be linearized as was done in ref.\ \cite{Floerchinger:2013rya}. When a Fourier expansion is used for the azimuthal dependence one has to deal with non-trivial boundary conditions for the radial dependence at $r=0$ and it is useful to employ again a variant of the Bessel expansion for the solution. Using a mode expansion similar to the one above also for the fluid velocity and shear stress leads to a set of coupled ordinary differential equations for the expansion coefficients. A truncated set of these can be easily solved numerically. 

One can actually also recover non-linear interactions between modes from this setup. Indeed, the equations of motion for the fluctuations have quadratic and higher order terms that can be integrated using the linear solutions as lowest order. One can in principle go on order by order in this perturbative scheme. Although this kind of perturbation theory has not been fully developed yet, scaling investigations of full numerical solutions indicate that it is well behaved and that convergence is rather good \cite{Floerchinger:2013tya}. For the rest of this proceedings contribution we concentrate on the leading linear term, however. 

The fluid dynamic equations of motion can be solved for the different modes labeled by the azimuthal wavenumber $m$ and the radial wavenumber $l$ and the solution can be stored. In principle one can do the same not only for initial density perturbations but also fluid velocity, shear stress etc. With these solutions at hand one can now come to the third point in the above program, the freeze out to particle spectra.

In close analogy to the fluid dynamic evolution one can use a background-fluctuation splitting also at kinetic freeze-out. The main contribution to the particle spectrum comes from the background part of the hydrodynamic fields and the former inherits the symmetries of the latter, in particular azimuthal rotation invariance. Since the hydrodynamic fields (fluid velocity, temperature) enter the particle spectrum in an essentially exponential way it is actually advantageous to expand its logarithm. One writes for a single event
\begin{equation}
\ln \left( \frac{dN^\text{single event}}{p_T dp_T d\phi dy} \right)
= {\underbrace{ \ln S_0(p_T)}_{\text{from background}}} + {\underbrace{\sum_{m=-\infty}^\infty \sum_{l=1}^\infty \; \tilde{w}^{(m)}_l \; e^{i m \phi} \; \theta^{(m)}_l(p_T).}_{\text{from fluctuations}}}
\end{equation}
Here $\tilde w^{(m)}_l$ are the weights of the initial transverse enthalpy distribution as in eq.\  \eqref{eq:BesselFourierExpDensity}. Note that these are complex (they carry information about a magnitude and an angle),
\begin{equation}
\tilde w^{(m)}_l = |\tilde w^{(m)}_l|\, e^{im\psi^{(m)}_l}.
\end{equation}
The functions $\theta^{(m)}_l(p_T)$ determine the $p_T$-dependence of the contribution of a particular mode. They are determined by evolving a particular mode hydrodynamically and determining its contribution to the particle spectrum at freeze.  The details of the kinetic freeze-out description in a background-fluctuation splitting approach have been described in ref.\ \cite{Floerchinger:2013hza}. 

One can now use the whole formalism not only for single events but also for event-by-event correlations or distributions. For example, the double differential harmonic flow coefficient is given to lowest order in the weights of initial density perturbations by the following intuitive matrix expression
\begin{equation}
v_m^2\{2\} (p_{T}^a, p_{T}^b) = \sum_{l_1,l_2=1}^{l_\text{max}} \theta^{(m)}_{l_1}(p_{T}^a)\; \theta^{(m)}_{l_2}(p_{T}^b) \;\left\langle \tilde w^{(m)}_{l_1} \tilde w^{(m)*}_{l_2} \right\rangle.
\label{eq:doublediffharmflow}
\end{equation}
It is interesting to note that this does in general not factorize into a product of two (identical) functions of $p_T^a$ and $p_T^b$. One can calculate the single-differential harmonic flow coefficients by using the usual definitions in terms of weighted $p_T$ integrals.

\begin{figure}
\begin{center}
\includegraphics[width=0.5\textwidth]{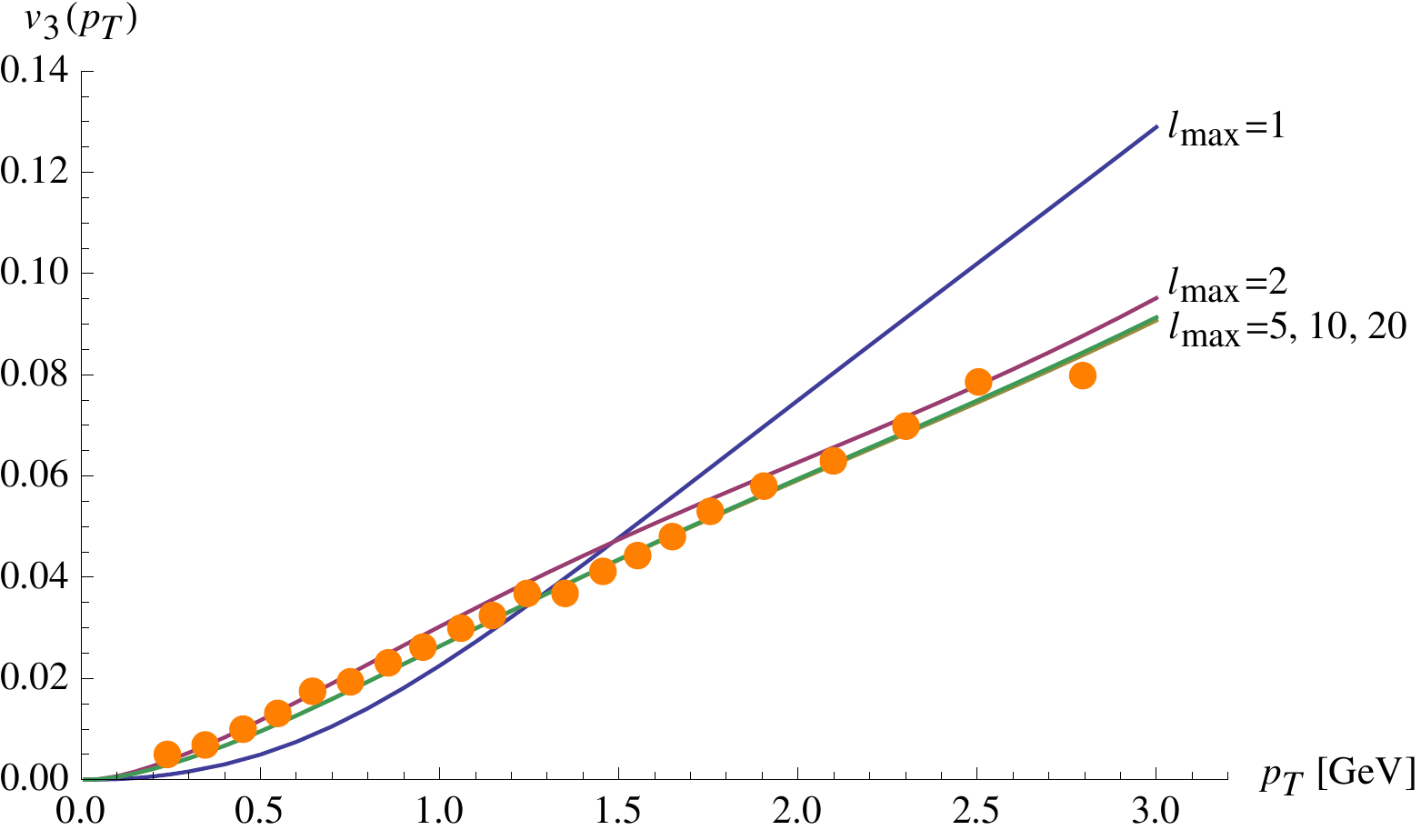}
\caption{Triangular flow of charged particles $v_3(p_T)$. Results from eq.\ \eqref{eq:doublediffharmflow} for different values of $l_\text{max}$, calculated by summing the contributions from  $\pi^\pm, K^\pm, p$ and $\bar p$ are compared to to experimental data on $2\%$ most central Pb+Pb collisions at $\sqrt{s_{\rm NN}} = 2.76$ TeV by ALICE \cite{ALICE:2011ab}. The figure is taken from ref.\ \cite{Floerchinger:2013rya}.}
\label{fig1}
\end{center}
\end{figure}

For central collisions the whole program has been carried out for initial conditions obtained from a Monte-Carlo-Glauber model in ref.\ \cite{Floerchinger:2013rya}. The $p_T$-differential harmonic flow coefficients $v_2$, $v_3$, $v_4$ and $v_5$ for charged particles agree reasonably well with experimental results obtained by the ALICE collaboration \cite{ALICE:2011ab}. In Fig.\ \ref{fig1} this is shown for triangular flow $v_3(p_T)$. It is interesting to compare different values for the maximum number of radial modes $l_\text{max}$ in eq.\ \eqref{eq:doublediffharmflow} and we do so in Fig.\ \ref{fig1}. The convergence is rather quick with the dominant part coming from the lowest modes $l=1$ and $l=2$. This means in turn that only a few initial modes can actually be efficiently constrained experimentally.

In conclusion, we have developed a semi-analytic method to characterize and propagate initial fluctuations in hydrodynamical fields. This has the advantage that the hydrodynamic equations can be solved independent of a concrete model for the initial state. A first application of the formalism was for initial enthalpy density fluctuations as calculated from a Monte-Carlo Glauber model and it lead to good agreement with experimental data for central collisions. From the numerical results it seems that fluctuation modes up to $l_\text{max} \approx 5$ can be resolved. However, this is expected to depend sensitively on the thermodynamic and transport properties which should be investigated more closely in the future.

In due course of this investigation it became apparent that it would also be interesting to extend the mode-by-mode treatment beyond linear order. In fact, a perturbative expansion treating the evolution of modes and their interaction order-by-order seems to be well behaved and quickly convergent \cite{Floerchinger:2013tya} and could provide interesting new insights. 

The main prospect for the mode-by-mode treatment on a long term might be a more detailed insight into the nature of the relativistic hydrodynamic evolution on the length and time scales of heavy ion collisions. This should show in particular how much ``discriminating power'' fluid dynamic studies have with respect to different initial conditions and QCD material properties and help to fully explore them. To that end it would be interesting to develop in more detail the non-linear mode-by-mode perturbation theory as outlined in \cite{Floerchinger:2013tya} and to extend the discussion from initial enthalpy density fluctuations in the transverse plane to more general initial fluctuations in hydrodynamic fields, in particular fluctuations in the fluid velocity, shear stress and baryon number density and to rapidity-dependent fluctuations.

\paragraph{Acknowledgements}
I thank Urs A. Wiedemann for collaboration on the topics discussed here and the organizers of IS 2013 for the nice conference.

\end{document}